\documentclass[jkps,twocolumn,fleqn,showpacs,showkeys]{revtex4}
\usepackage{graphicx,amssymb,amsmath,bm}
\newtheorem{thm}{Theorem}

\begin{document}
\title[]{Time Delay Effect on the Love Dynamical Model}
\author{Woo-Sik \surname{Son}}
\email{woosik.son@gmail.com}
\thanks{Fax: +82-2-701-7427}
\author{Young-Jai \surname{Park}}
\affiliation{Department of Physics,
Sogang University, Seoul 121-742, Korea}
\affiliation{Department of
Service Systems Management and Engineering, Sogang University, Seoul 121-742, Korea}
\date[]{}

\begin{abstract}

We investigate the effect of time delay on the dynamical model of love.
The local stability analysis proves that the time delay on the return
function can cause a Hopf bifurcation and a cyclic love dynamics.
The condition for the occurrence of the Hopf bifurcation is also
clarified. Through a numerical bifurcation analysis, we confirm the
theoretical predictions on the Hopf bifurcation and obtain a universal
bifurcation structure consisting of a supercritical Hopf bifurcation and a
cascade of period-doubling bifurcations, i.e., a period doubling
route to chaos.

\end{abstract}


\maketitle

\section{INTRODUCTION}
\label{sec:1}

In a one page pioneering paper \cite{Strogatz} and a book
\cite{Strogatz1}, Strogatz suggested a simple pedagogical model
describing a love affair. His goal was to teach harmonic oscillation
phenomena using ``a topic that is already on the minds of many
college students: the time-evolution of a love affair between two
people''. Later, Rinaldi, Gragnani, and Feichtinger proposed more
realistic mathematical models for love dynamics
\cite{Rinaldi,Rinaldi1,Gragnani,Rinaldi2}: They showed that dynamic
phenomena in the field of social science can also be analyzed by using a 
modeling approach via ordinary differential equations. Their models
explained that two individuals, who are completely indifferent to
each other from the start, approach a plateau of love affair. They
also showed that the coexistence of insecurity and synergism results
in a cyclic dynamics of romantic feelings. Secure individuals react
positively to their partner's love and are not afraid about their
partner becoming emotionally close to them while non-secure ones
react negatively to high involvement \cite{Rinaldi1}. Synergic
individuals are those who increase their reactions to their partner's
appeal when they are in love \cite{Gragnani}. Thereafter, following
a suggestion of Strogatz, Sprott investigated the dynamics of a love
triangle, which produce a chaotic behavior \cite{Sprott}, and 
suggested dynamical models of happiness \cite{Sprott1}. Moreover,
Wauer et al. studied the love dynamics for time-varying fluctuations
\cite{Wauer}. Very recently, Rinaldi et al. constructed 
full catalog of possible love stories among two individuals
\cite{Rinaldi3}, and Barley and Cherif studied the stochastic
love dynamical model \cite{Barley}.

On the other hand, a dynamical model described by using delay
differential equations (DDEs) has attracted much attention in
various fields of science, e.g., biology \cite{Mackey,Wei},
chemistry \cite{Epstein}, neural systems \cite{Kook,Xu}, excitable systems
\cite{Sethia}, transport control \cite{Son2,Son1} and cryptography
\cite{Kye,Kye1}. DDEs also support a realistic mathematical modeling
of economic dynamics \cite{Cesare,Neamtu,Son}. In the field of
social science, Liao and Ran recently showed that the time delay on
love dynamics can cause a Hopf bifurcation \cite{Liao}. However,
they did not consider the linear, secure, and non-secure returns.
Also, they did not deal with the synergic instinct. For that reason,
they could not clarify the condition for the occurrence of Hopf
bifurcation.

In this paper, we investigate the effect of time delay on the
nonlinear dynamical model describing a love affair between two
individuals. By analyzing the characteristic equation of
linearization of the model, we theoretically prove that if no one
exhibits a non-secure return in both cases of synergic and
non-synergic couples, then the existence of time delay cannot
disturb a plateau of love affair, i.e., steady state. However, if at
least one of them has a non-secure return, then the time delay on
return function can cause a Hopf bifurcation and a cyclic love
dynamics. On the other hand, through numerical bifurcation analysis,
we confirm the theoretical results on Hopf bifurcation and
investigate additional bifurcation phenomena. As a result, we obtain
a universal bifurcation structure consisting of a supercritical Hopf
and a cascade of period-doubling bifurcations, resulting in chaotic
motion for our models.

This paper is organized as follows: Section \ref{sec:2} presents the
love dynamical model. In Section \ref{sec:3}, we prove the
occurrence of Hopf bifurcation by using a local stability analysis. In
Section \ref{sec:4}, we show the results of our numerical bifurcation
analysis. The conclusion is given in Section \ref{sec:5}.

\section{Love dynamical model}
\label{sec:2}

Now, let us investigate the love dynamical model based on a series
of models proposed by Rinaldi, Gragnani, and Feichtinger
\cite{Rinaldi,Rinaldi1,Gragnani}. The model has a variable $x_i$
($i=1,2$), which is a measure of the love of an individual $i$ for
his or her partner $j$ ($j=2,1$). Positive values of $x_i$ represent 
love while negative values are associated with hate. Complete
indifference is identified by $x_i = 0$.

It is important to mention the time scale of the love dynamical model.
Fast fluctuations of the feelings influenced by daily or weekly
activities cannot be captured by the model. Also, the learning and
the adaptation processes over a long range of time are not considered.
Thus, the following model can only be used on an intermediate time
scale (months/years), for example, in predicting if a love story will
be characterized by stationary or stormy feelings \cite{Rinaldi1}.

The dynamics of love is comprised of three basic processes: oblivion
$O_i$, return $R_i$ and instinct $I_i$.

\begin{equation}\label{eq:1}
\dot{x}_i (t) = O_i\big(x_{i}(t)\big) + R_{i}\big(x_{j}(t)\big) +
I_{i}\big(x_{i}(t)\big)
\end{equation}

\noindent In the following, $x_i (t)$ and $x_j (t)$ are replaced
by $x_i$ and $x_j$ for compact notation. Oblivion is described
by

\begin{equation}\label{eq:2}
O_{i}(x_{i}) = -\alpha_i x_{i},
\end{equation}

\noindent where $\alpha_i >0$ is a forgetting coefficient.
Therefore, $x_i$ decays exponentially when an individual $i$ loses
partner $j$ ($R_i = I_i = 0$).

The return $R_i$ is related to the reaction of individual $i$ to the
partner's love $x_j$. Three different types of return functions have
been considered, namely, the linear return ${R_i}^l$, the secure return
${R_i}^s$, and the non-secure return ${R_i}^n$. The simplest one is the
linear return \cite{Rinaldi} given by

\begin{equation}\label{eq:3}
{R_i}^{l}(x_{j}) = \beta_i x_j,
\end{equation}

\noindent where $\beta_i > 0$ is a reactiveness to the love. It is
unbounded and describes that an individual $i$ ``loves to be loved''
and ``hates to be hated''. The secure return \cite{Rinaldi1} is
specified by

\begin{eqnarray}\label{eq:4}
{R_i}^{s}(x_{j}) = \left\{ \begin{array}{ll} \beta_i x_j/(1+x_j ) & \textrm{for \,$x_j \ge 0$}, \\
                           \beta_i x_j/(1-x_j ) & \textrm{for \,$x_j <
                           0$}.
                           \end{array} \right.
\end{eqnarray}

\begin{figure}
\begin{center}
\includegraphics[width=9.5cm]{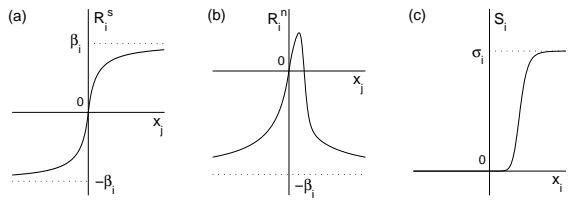}
\caption{Shape of the functions: (a) secure return ${R_i}^s
(x_{j})$, (b) non-secure return ${R_i}^n (x_{j})$, and (c) synergic
function $S_{i}(x_{i})$.}\label{fig:fig1.eps}
\end{center}
\end{figure}

\noindent It is an increasing and bounded function, as shown in Fig.
\ref{fig:fig1.eps}(a). Secure individuals react positively to
their partner's love and are not afraid about their partner
becoming emotionally close to them. However, non-secure
individuals react negatively to high pressures and involvement, as
shown in Fig. \ref{fig:fig1.eps}(b). The non-secure return
\cite{Gragnani} is described by

\begin{eqnarray}\label{eq:5}
{R_i}^{n}(x_{j}) = \left\{ \begin{array}{ll} \beta_i x_j
(1-{x_{j}}^{8}) /\{(1+x_j ) (1+{x_{j}}^{8})\} \\
                           \qquad \qquad \qquad \qquad \qquad \quad \ \ \: \textrm{for \,$x_j \ge 0$}, \\
                           \beta_i x_j/(1-x_j ) \qquad \qquad \qquad \textrm{for \,$x_j <
                           0$}.
                           \end{array} \right.
\end{eqnarray}

The instinct $I_i$ is related to the reaction of individual $i$ to
the partner's appeal $A_j$. In the following, we only consider
positive appeal $A_j$. Two different types of instinct functions
have been suggested, namely the synergic instinct ${I_i}^s$ and
non-synergic instinct ${I_i}^n$. The non-synergic instinct
\cite{Rinaldi,Rinaldi1} is given by

\begin{equation}\label{eq:6}
{I_i}^n = \gamma_i A_j,
\end{equation}

\noindent where $\gamma_i > 0$ is a reactiveness to the appeal. When
the synergic instinct is considered, the individual's reaction to
the partner's appeal can be enhanced by love. For example, mothers
often have a biased view of the beauty of their children. The
synergic instinct \cite{Gragnani} is described by

\begin{eqnarray}\label{eq:7}
{I_i}^s&=&\{ 1+S_{i}(x_{i}) \}\,\gamma_i A_{j}, \\
S_{i}(x_{i})&=&\left\{ \begin{array}{ll} \sigma_i
{x_i}^8/(1+{x_i}^8) & \textrm{for \,$x_i \ge 0$}, \\
                           0 & \textrm{for \,$x_i <
                           0$},
                           \end{array} \right. \nonumber
\end{eqnarray}

\noindent where the synergic function $S_{i}(x_{i})$ is an
increasing and bounded function for $x_{i} \ge 0$, as shown in Fig.
\ref{fig:fig1.eps}(c).

Now, we can take the essential step for describing our love
dynamical model. How does an individual know the partner's romantic
feeling? In a real situation, the romantic interaction is mediated
by communication, e.g., a talk, a phone call, an email, a letter,
etc. That is, time is required for the romantic feelings of
someone to transfer to the other. In addition, Ackerman et al. very
recently showed that an individual could delay one's confessing
love to adjust potential costs and benefits \cite{Ackerman}.

Therefore, the oblivion, the return, and the instinct in the model
of Eq. (\ref{eq:1}) do not proceed {\it simultaneously}, and the delay
time $\tau$ must be added to the return function. As a result, the
love dynamical model can be described by DDEs as follows:

\begin{equation}\label{eq:8}
\dot{x}_i (t) = O_i\big(x_{i}(t)\big) + R_{i}\big(x_{j}(t-\tau)\big)
+ I_{i}\big(x_{i}(t)\big).
\end{equation}

\noindent For the purpose of simplification, we consider the same
delay time $\tau$ for both individuals.

It seems appropriate to comment on the recent work of Liao and Ran
\cite{Liao}. They introduced the time delay on return functions and
observed the occurrence of a Hopf bifurcation. Using the same notation
as Ref. \cite{Liao}, their model is represented by

\begin{equation}\label{eq:8-1}
\dot{x}_i (t) =
-a_{i}x_{i}(t)+b_{i}f(x_{j}(t-\tau_{j}))+\gamma_{i}A_{j}.
\end{equation}

\noindent It contains the same basic processes of love dynamics:
oblivion, return, and instinct. However, their return function
$b_{i}f(x_{j}(t-\tau_{j}))$ could not fully consider the linear,
the secure, and the non-secure return functions. Also, they did not deal with
the synergic instinct. As a result, they could not clarify in which
type of return functions the Hopf bifurcation can arise.

In their model, the condition for a Hopf bifurcation results in

\begin{equation}\label{eq:8-2}
0>-a_{1}a_{2}>b_{1}b_{2}\left.\frac{df}{dx_{1}}\right|_{x_{1}^{*}}
\left.\frac{df}{dx_{2}}\right|_{x_{2}^{*}},
\end{equation}

\noindent where $(x_{1}^{*},x_{2}^{*})$ is a fixed point of the
model in Eq. (\ref{eq:8-1}). They also showed a numerical example of a Hopf
bifurcation in which $a_{1}=a_{2}=1$, $b_{1}=1.5$, $b_{2}=-2$, and
$f(x)=\tanh(x)$. Note that $\tanh(x)$ is an increasing and bounded
function, so individuals 1 and 2 exhibit a secure $(b_{1}>0)$ and
an {\it anti-secure} return $(b_{2}<0)$, respectively. 
But an {\it anti-secure} individual seems unusual.

Therefore, we will fully investigate the effect of time delay on the
love dynamical models suggested by Rinaldi, Gragnani, and
Feichtinger \cite{Rinaldi,Rinaldi1,Gragnani}. In the following
section, we will explicitly show that, in both cases of synergic and
non-synergic couples, if at least one individual exhibits a
non-secure return, then the time delay on the return function can cause
a Hopf bifurcation.

\section{Hopf bifurcation analysis}
\label{sec:3}

\subsection{Non-synergic Couple}\label{sec:3-1}

First, let us consider a couple composed of non-synergic
individuals. We follow the same steps in Refs. \cite{Wei,Son} for
verifying the occurrence of a Hopf bifurcation. The love dynamics for
the non-synergic couple is described by

\begin{eqnarray}\label{eq:9}
\dot{x_1}(t) &=& -\alpha_1 x_{1}(t) +
R_{1}\big(x_{2}(t-\tau)\big)+\gamma_1
A_{2}, \\
\dot{x_2}(t) &=& -\alpha_2 x_{2}(t) +
R_{2}\big(x_{1}(t-\tau)\big)+\gamma_2 A_{1}. \nonumber
\end{eqnarray}

\noindent Here, we do not need to fix the type of return function
$R_i$ at this stage. It may be one of the linear, secure, and
non-secure returns. We assume that $(x_1 ^{*},x_2 ^{*})$ is a fixed
point of the model in Eq. (\ref{eq:9}), which is located in the first
quadrant. Then, the linearization of the model in Eq. (\ref{eq:9}) at
$(x_1^{*},x_2^{*})$ is given by

\begin{eqnarray}\label{eq:10}
\dot{\delta x_1}(t)&=& -\alpha_{1}\delta
{x_1}(t)+c_{1}\delta x_2 (t-\tau), \\
\dot{\delta x_2}(t)&=& -\alpha_{2}\delta {x_2}(t)+c_{2}\delta x_1
(t-\tau), \nonumber
\end{eqnarray}

\noindent where $c_{1} = dR_{1}/dx_{2}|_{x_2 ^*}$ and $c_{2} =
dR_{2}/dx_{1}|_{x_1 ^*}$. The characteristic equation of
Eq. (\ref{eq:10}) is described by

\begin{equation}\label{eq:11}
\lambda^2 +(\alpha_1 + \alpha_2 )\lambda + \alpha_1 \alpha_2 - c_1
c_2 e^{-2 \lambda \tau} = 0.
\end{equation}

\noindent When $\tau =0$, all roots of Eq. (\ref{eq:11}) have real
negative parts if and only if the conditions

\begin{eqnarray}
(\boldsymbol{\mathrm{H_{1}}}) \ \, \alpha_1 + \alpha_2 > 0, \ \,
\alpha_1 \alpha_2 -c_1 c_2 > 0 \nonumber
\end{eqnarray}

\noindent hold.

Now, let us assume that $i\omega$ (real positive $\omega$) is a root
of Eq. (\ref{eq:11}). Then, we have

\begin{eqnarray}\label{eq:12}
\omega^2 - \alpha_1 \alpha_2 &=& -c_1 c_2 \cos 2\omega \tau, \\
(\alpha_1 + \alpha_2 ) \omega &=& -c_1 c_2 \sin 2\omega \tau,
\nonumber
\end{eqnarray}

\noindent which lead to

\begin{equation}\label{eq:13}
p^2 + ({\alpha_1}^{2} +{\alpha_2}^{2} ) p + {\alpha_1}^{2}{\alpha_2}
^{2}-{c_1}^{2}{c_2}^{2} = 0,
\end{equation}

\noindent where $p=\omega ^2$. It follows that if the conditions

\begin{eqnarray}
(\boldsymbol{\mathrm{H_{2}}}) \ \,  {\alpha_1}^{2} +{\alpha_2}^{2} >
0, \ \, {\alpha_1}^{2}{\alpha_2}^{2} -{c_1}^{2}{c_2}^{2} > 0
\nonumber
\end{eqnarray}

\noindent are satisfied, then Eq. (\ref{eq:13}) has no positive
roots. Hence, all roots of Eq. (\ref{eq:11}) have real negative
parts when $\tau \in [0,\infty)$.

On the other hand, if the condition

\begin{eqnarray}
(\boldsymbol{\mathrm{H_{3}}}) \ \,  {\alpha_1}^{2}{\alpha_2}^{2}
-{c_1}^{2}{c_2}^{2} < 0 \nonumber
\end{eqnarray}

\noindent holds, then Eq. (\ref{eq:13}) has a unique positive root
$p_0 = {\omega_0}^2$. Substituting $\omega_0 $ into Eq.
(\ref{eq:12}), we have the results

\begin{eqnarray}\label{eq:14}
\omega_0 &=& \frac{1}{\sqrt{2}}\Big[ \sqrt{ ({\alpha_1}^{2}
-{\alpha_2} ^{2})^2
+ 4{c_1}^{2}{c_2}^{2} }-({\alpha_1}^{2} +{\alpha_2}^{2})  \Big]^{1/2}, \nonumber \\
\tau_n &=& \frac{1}{2\omega_0}\cos^{-1}\Big[ \frac{\alpha_1 \alpha_2
-{\omega_0}^2}{c_1 c_2} \Big] + \frac{n\pi}{\omega_0},
\end{eqnarray}

\noindent for $n=0,1,2,\cdots$. Then, let us investigate the sign of
$\mathrm{Re}\,[d\lambda/d\tau]$. The differentiation of Eq.
(\ref{eq:11}) with respect to $\tau$ and the substitutions $\tau=\tau_0$
and $\lambda = i \omega_0$ lead to

\begin{equation}\label{eq:15}
\textrm{Re}\Big[\,\frac{d\lambda}{d\tau}\,\Big]_{
\tau=\tau_{0},\omega=\omega_{0}}  =\frac{4{\omega_0} ^4 +
2{\omega_0} ^2 ({\alpha_1} ^2 + {\alpha_2} ^2)}{A^2+B^2},
\end{equation}

\noindent where

\begin{eqnarray}
A &=& (\alpha_1 + \alpha_2 )\cos 2\omega_0 \tau_0 -2\omega_0 \sin
2\omega_0 \tau_0 +2c_1 c_2 \tau_0 , \nonumber \\
B &=& (\alpha_1 + \alpha_2 )\sin 2\omega_0 \tau_0 +2\omega_0 \cos
2\omega_0 \tau_0 . \nonumber
\end{eqnarray}

\noindent It is easily obtained that

\begin{equation}\label{eq:16}
\textrm{Re}\Big[\,\frac{d\lambda}{d\tau}\,\Big]_{
\tau=\tau_{0},\omega=\omega_{0}} > 0.
\end{equation}

As a result, we have the following theorem from Corollary 2.4 in
Ref. \cite{Wei}:

\begin{thm}
{~}
\begin{enumerate}
\item If \textup{($\boldsymbol{\mathrm{H_{1}}}$)} and
\textup{($\boldsymbol{\mathrm{H_{2}}}$)} hold, then the fixed point
$(x_1^{*},x_2^{*})$ is asymptotically stable for all $\tau \geq 0$.
\item If \textup{($\boldsymbol{\mathrm{H_{1}}}$)} and
\textup{($\boldsymbol{\mathrm{H_{3}}}$)} hold, then the fixed point
$(x_1^{*},x_2^{*})$ is asymptotically stable for $\tau < \tau_{0}$
and unstable for $\tau > \tau_{0}$. Furthermore,
 the love dynamical model in Eq. \textup{(\ref{eq:9})} undergoes a Hopf bifurcation at
$(x_1^{*},x_2^{*})$ when $\tau =\tau_{0}$.
\end{enumerate}
\end{thm}

\noindent Note that the linear return ${R_i}^{l}(x_{j})$ and the secure
return ${R_i}^{s}(x_{j})$ are always increasing functions with $c_i
> 0$, in contrast to the non-secure return. Because $\alpha_1$ and $\alpha_2$ are positive, the
conditions ($\boldsymbol{\mathrm{H_{1}}}$) and
($\boldsymbol{\mathrm{H_{3}}}$) result in

\begin{equation}\label{eq:17}
0 > -\alpha_1 \alpha_2 > c_1 c_2 .
\end{equation}

\noindent Therefore, if no one exhibits a non-secure return, then
the existence of time delay cannot disturb a steady state
$(x_1^{*},x_2^{*})$. However, if at least one of them has a
non-secure return and the inequality in Eq. (\ref{eq:17}) is satisfied,
then the time delay on the return function leads to a Hopf bifurcation
and a cyclic love dynamics.

\subsection{Synergic Couple}

Second, let us investigate the effect of time delay on a couple
composed of synergic individuals. In this case, the love dynamics is
given by

\begin{eqnarray}\label{eq:18}
\dot{x_1}(t)=-\alpha_1 x_{1}(t)&+&
R_{1}\big(x_{2}(t-\tau)\big) \\
&+&\big\{ 1+S_{1}\big(x_{1}(t)\big) \big\}\,\gamma_1 A_{2}, \nonumber \\
\dot{x_2}(t)=-\alpha_2 x_{2}(t)&+&
R_{2}\big(x_{1}(t-\tau)\big) \nonumber \\
&+&\big\{ 1+S_{2}\big(x_{2}(t)\big) \big\}\,\gamma_2 A_{1}.
\nonumber
\end{eqnarray}

\noindent Also, the type of return function $R_i$ is not fixed at
this stage. It may be one of the linear, secure, and non-secure
returns. Let $(x_1 ^{*},x_2 ^{*})$ be a fixed point of the model
in Eq. (\ref{eq:18}), which is located in the first quadrant. The
linearization of the model in Eq. (\ref{eq:18}) at $(x_1 ^{*},x_2 ^{*})$ is
given by

\begin{eqnarray}\label{eq:19}
\dot{\delta x_1}(t)&=& -\alpha_{1}\delta
{x_1}(t)+c_{1}\delta x_2 (t-\tau) + d_{1}\delta{x_1}(t), \\
\dot{\delta x_2}(t)&=& -\alpha_{2}\delta {x_2}(t)+c_{2}\delta x_1
(t-\tau) + d_{2}\delta{x_2}(t), \nonumber
\end{eqnarray}

\noindent where $d_{1}= dS_{1}/dx_{1}|_{x_1 ^*} \cdot \gamma_1 A_2$,
$d_{2}= dS_{2}/dx_{2}|_{x_2 ^*} \cdot \gamma_2 A_1$. Then, the
characteristic equation of Eq. (\ref{eq:19}) is described by

\begin{equation}\label{eq:20}
\lambda^2 +( \alpha_1 ^s + \alpha_2 ^s ) \lambda + \alpha_1 ^s
\alpha_2 ^s - c_1 c_2 e^{-2 \lambda \tau} = 0,
\end{equation}

\noindent where $\alpha_i ^s = \alpha_i -d_i$. For the case of $\tau
= 0$, all roots of Eq. (\ref{eq:20}) have real negative parts if and
only if the conditions

\begin{eqnarray}
(\boldsymbol{\mathrm{H_{4}}}) \ \, \alpha_1 ^s + \alpha_2 ^s
> 0, \ \, \alpha_1 ^s \alpha_2 ^s - c_1 c_2 > 0 \nonumber
\end{eqnarray}

\noindent are satisfied. Then, assuming that $i\omega$ (real positive
$\omega$) is a root of Eq. (\ref{eq:20}), we obtain

\begin{equation}\label{eq:21}
p^2 + \{(\alpha_1 ^s)^2 + (\alpha_2 ^s)^2 \}p + (\alpha_1 ^s )^2
(\alpha_2 ^s )^2 -{c_1}^2 {c_2}^2 = 0.
\end{equation}

\noindent where $p=\omega^2$. In the same manner as in Section
\ref{sec:3-1}, it follows that if the conditions

\begin{eqnarray}
(\boldsymbol{\mathrm{H_{5}}}) \ \,  (\alpha_1 ^s )^{2} +(\alpha_2 ^s
)^{2} > 0, \ \, (\alpha_1 ^s )^{2}(\alpha_2 ^s )^{2}
-{c_1}^{2}{c_2}^{2}
> 0 \nonumber
\end{eqnarray}

\noindent hold, then all roots of Eq. (\ref{eq:20}) have real
negative parts when $\tau \in [0,\infty)$. If the condition

\begin{eqnarray}
(\boldsymbol{\mathrm{H_{6}}}) \ \, (\alpha_1 ^s )^2 (\alpha_2 ^s )^2
-{c_1}^2 {c_2}^2 < 0 \nonumber
\end{eqnarray}

\noindent is satisfied, then Eq. (\ref{eq:21}) has a unique positive
root $p_0 = {\omega_0}^2$. Accordingly, we have the results

\begin{eqnarray}\label{eq:21-1}
\omega_0&=&\frac{1}{\sqrt{2}}\Big[ \sqrt{ \big\{(\alpha_1 ^s )^{2}
-(\alpha_2 ^s ) ^{2}\big\}^2
+ 4{c_1}^{2}{c_2}^{2} } \\
{}&& -\big\{ (\alpha_1 ^s )^{2} +(\alpha_2 ^s )^{2} \big\}  \Big]^{1/2}, \nonumber \\
\tau_n &=&\frac{1}{2\omega_0}\cos^{-1}\Big[ \frac{\alpha_1 ^s
\alpha_2 ^s -{\omega_0}^2}{c_1 c_2} \Big] + \frac{n\pi}{\omega_0},
\nonumber
\end{eqnarray}

\noindent for $n=0,1,2,\cdots$.

Now, the remaining step is to fix the sign of
$\mathrm{Re}\,[d\lambda/d\tau]$. We can obtain

\begin{equation}\label{eq:21-2}
\textrm{Re}\Big[\,\frac{d\lambda}{d\tau}\,\Big]_{
\tau=\tau_{0},\omega=\omega_{0}}  =\frac{4{\omega_0} ^4 +
2{\omega_0} ^2 \big\{(\alpha_1 ^s ) ^2 + (\alpha_2 ^s )
^2\big\}}{C^2+D^2},
\end{equation}

\noindent where

\begin{eqnarray}
C &=& (\alpha_1 ^s  + \alpha_2 ^s )\cos 2\omega_0 \tau_0 -2\omega_0
\sin
2\omega_0 \tau_0 +2c_1 c_2 \tau_0 , \nonumber \\
D &=& (\alpha_1 ^s + \alpha_2 ^s )\sin 2\omega_0 \tau_0 +2\omega_0
\cos 2\omega_0 \tau_0 . \nonumber
\end{eqnarray}

\noindent This leads to

\begin{equation}\label{eq:21-3}
\textrm{Re}\Big[\,\frac{d\lambda}{d\tau}\,\Big]_{
\tau=\tau_{0},\omega=\omega_{0}} > 0.
\end{equation}

As a result, we have the following theorem from Corollary 2.4 in
Ref. \cite{Wei}:

\begin{thm}
{~}
\begin{enumerate}
\item If \textup{($\boldsymbol{\mathrm{H_{4}}}$)} and
\textup{($\boldsymbol{\mathrm{H_{5}}}$)} hold, then the fixed point
$(x_1^{*},x_2^{*})$ is asymptotically stable for all $\tau \geq 0$.
\item If \textup{($\boldsymbol{\mathrm{H_{4}}}$)} and
\textup{($\boldsymbol{\mathrm{H_{6}}}$)} hold, then the fixed point
$(x_1^{*},x_2^{*})$ is asymptotically stable for $\tau < \tau_{0}$
and unstable for $\tau > \tau_{0}$. Furthermore,
 the love dynamical model in Eq. \textup{(\ref{eq:9})} undergoes a Hopf bifurcation at
$(x_1^{*},x_2^{*})$ when $\tau =\tau_{0}$.
\end{enumerate}
\end{thm}

\noindent Note that the conditions ($\boldsymbol{\mathrm{H_{4}}}$)
and ($\boldsymbol{\mathrm{H_{6}}}$) result in

\begin{equation}\label{eq:22}
0 > -|\alpha_1 ^s \alpha_2 ^s | > c_1 c_2 .
\end{equation}

\noindent Thus, if no one exhibits a non-secure return, then the
time delay cannot destabilize a steady state $(x_1^{*},x_2^{*})$. On
the other hand, if at least one of them has a non-secure return and
the inequality in Eq. (\ref{eq:22}) is satisfied, then the time delay on
the return function leads to a Hopf bifurcation and a cyclic love
dynamics. Therefore, we obtain the same results as in the non-synergic
case.

\section{Numerical bifurcation analysis}
\label{sec:4}

In the previous section, we have theoretically proven the occurrence
of a Hopf bifurcation. In this section, we show that a numerical
bifurcation analysis supports the theoretical results on Hopf
bifurcation and investigate additional bifurcation phenomena. The
reactiveness to the love, i.e., $\beta_i$, and the delay time $\tau$ are
considered as varying parameters, and the others are fixed. For
numerical detection and continuation of a bifurcation point in
parameter space $(\beta_i ,\tau)$, we use DDE-BIFTOOL
\cite{DDEBIFTOOL} and KNUT \cite{KNUT}.

\subsection{Non-synergic Couple}
\label{sec:4-1}

First, let us investigate a non-synergic couple. Among various
combinations, a couple composed of secure and non-secure
individuals are considered as a proper case for observing the Hopf
bifurcation based on the theoretical results. Thus, the love
dynamics is described by

\begin{eqnarray}\label{eq:23}
\dot{x_1}(t) &=& -\alpha_1 x_{1}(t) + {R_{1}}^s
\big(x_{2}(t-\tau)\big)+\gamma_1
A_{2}, \\
\dot{x_2}(t) &=& -\alpha_2 x_{2}(t) + {R_{2}}^n
\big(x_{1}(t-\tau)\big)+\gamma_2 A_{1}, \nonumber
\end{eqnarray}

\noindent where the functional forms of the secure return ${R_1}^s$
and the non-secure return ${R_2}^n$ are denoted in Eqs. (\ref{eq:4})
and (\ref{eq:5}), respectively.

\begin{figure}
\begin{center}
\includegraphics[width=8.8cm]{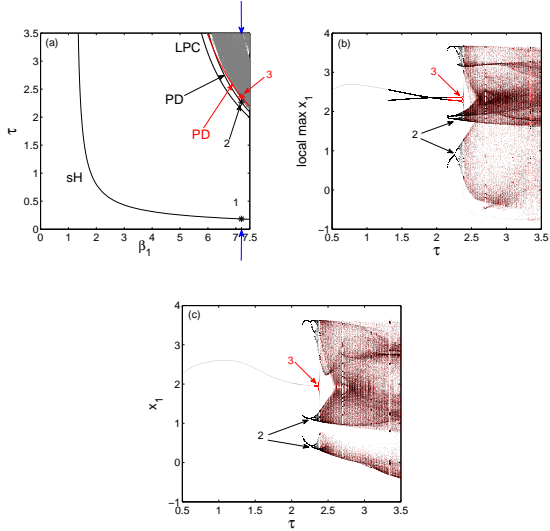}
\caption{(a) Bifurcation diagram in $(\beta_1 ,\tau)$ for the parameter
value: $\alpha_1 = \alpha_2 = 1$, $\beta_2 = 1$, and $\gamma_1 A_2
= \gamma_2 A_1 = 0.5$. (b) Orbit diagram of local maxima of $x_1$ at
a fixed value $\beta_{1}=7.2$. (c) Orbit diagram based on Poincare section ($x_2 = 0$).}\label{fig:fig2.eps}
\end{center}
\end{figure}

Figure \ref{fig:fig2.eps}(a) is a bifurcation diagram in $(\beta_1
,\tau)$. It shows a supercritical Hopf (sH), a limit point
bifurcation of cycles (LPC) and two period doubling (PD) bifurcation
curves. In this figure, the two PD curves represent the bifurcation from
period-1 to period-2 for two different branches of the limit cycle. The
black PD curve corresponds to the period doubling of the stable limit
cycle branch emerging from the LPC curve. On the other hand, the red
one is associated with the period doubling of the limit cycle branch
arising from the sH curve. Figure \ref{fig:fig2.eps}(a) also shows
a chaotic region represented by gray dots. That is determined by
the largest Lyapunov exponent $\lambda_{1}$ of the model
in Eq. (\ref{eq:23}). For this computation, we firstly obtain the time
series of DDEs from the constant history function
$x_{1}(t)=x_{2}(t)=0$ for $-\tau \leq t \leq 0$. Remember that we
assume the two individuals to be completely indifferent to each other when
they first meet. Then, the largest Lyapunov exponent of the time
series is estimated with the help of TISEAN \cite{TISEAN}. The
result of Fig. \ref{fig:fig2.eps}(a) also shows bistable
phenomena: Two limit cycles emerging from the sH and the LPC curves
coexist. Moreover, a limit cycle and a chaotic attractor coexist
around the period-doubling bifurcation point `2' in Fig.
\ref{fig:fig2.eps}(a). It shows that the red PD curve is on the
chaotic region.

In Fig. \ref{fig:fig2.eps}(b), we plot the orbit diagram, i.e., the
local maxima of the $x_{1}$ variable as varying $\tau$ at a fixed
$\beta_{1} = 7.2$. Here, the black and the red dots are obtained from $x_{1}(t)=x_{2}(t)=0$ and $x_{1}(t)=2.4,x_{2}(t)=0.2$ for $-\tau \leq t \leq 0$, respectively. 
The orbit diagram corresponds to the line indicated by blue
arrows in Fig. \ref{fig:fig2.eps}(a). In the following bifurcation
diagrams, lines indicated by blue arrows correspond to their
matching orbit diagrams. The result of Fig. \ref{fig:fig2.eps}(b) shows a cascade of 
period-doubling bifurcations. For a clearer illustration, we plot together the orbit diagram based on 
Poincare section ($x_2 = 0$) at the same fixed value $\beta_{1} = 7.2$. 
It explicitly shows a period-doubling route to chaos.
However, in the following orbit diagrams, we only plot those consisting of local
extrema, because they are clearer than those based on Poincare section
to understand how the bifurcation occurs.
Also, Fig. \ref{fig:fig2.eps}(b) and Fig. \ref{fig:fig2.eps}(c) agree well with the result
that the time delay can induce a period doubling route to chaos
\cite{Kitano}.

It seems appropriate to comment on more PD curves. In Fig.
\ref{fig:fig2.eps}(a), we could not present more PD curves
corresponding to the bifurcation from period-2 to period-4 or for
higher period because numerical detection and continuation of a
bifurcation point in DDEs is more subtle than that of ODEs. However,
the result of the orbit diagram in Fig. \ref{fig:fig2.eps}(b)
clearly explains that there exists a cascade of period doubling
bifurcations in Fig. \ref{fig:fig2.eps}(a). Concerning this diagram,
the non-zero history function does not agree with our real
experience and the assumption on initial indifference. However,
we add it to clarify the bistable phenomena observed in Fig.
\ref{fig:fig2.eps}(a). The orbit diagram explicitly shows 
two coexisting limit cycles and the coexistence of a limit cycle and a chaotic
attractor. The points `2' and `3' in Figs. \ref{fig:fig2.eps}(a) and
\ref{fig:fig2.eps}(b) represent the same points.

\begin{figure}
\begin{center}
\includegraphics[width=5cm]{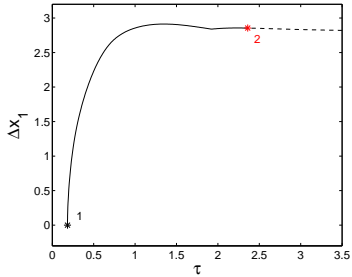}
\caption{Branch of the limit cycle emanating from the Hopf point and the
steady state for the parameter value: $\alpha_1 = \alpha_2 = 1$,
$\beta_1 = 7.2$, $\beta_2 = 1$, and $\gamma_1 A_2 = \gamma_2 A_1 =
0.5$. The points `1' and `2' represent the same points in Figs.
\ref{fig:fig2.eps}(a) and
\ref{fig:fig2.eps}(b).}\label{fig:fig3.eps}
\end{center}
\end{figure}

\begin{figure}
\begin{center}
\includegraphics[width=8.8cm]{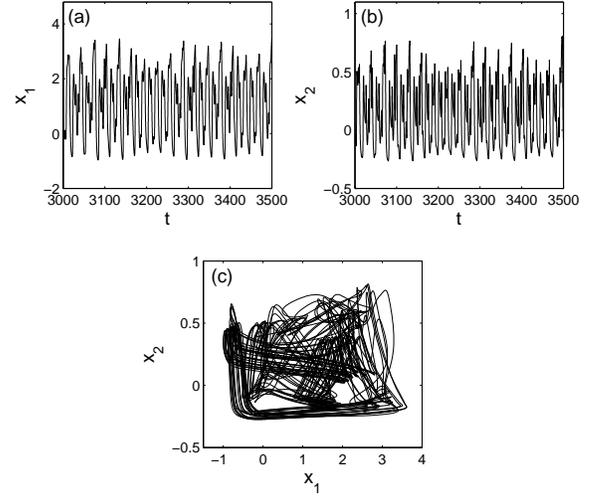}
\caption{(a,b) Time series and (c) phase plot of the model
in Eq. (\ref{eq:23}) for the parameter value: $\alpha_1 = \alpha_2 = 1$,
$\beta_1 = 7.2$, $\beta_2 = 1$, $\gamma_1 A_2 = \gamma_2 A_1 = 0.5$,
and $\tau = 3.2$. The results are obtained from
$x_{1}(t)=x_{2}(t)=0$ for $t \leq 0$.}\label{fig:fig4.eps}
\end{center}
\end{figure}

Now, let us show that the Hopf bifurcation in Fig.
\ref{fig:fig2.eps}(a) is supercritical. Though, for determining the
direction of Hopf bifurcation, rigorous analysis based on the normal
form method and the center manifold theory presented in Refs.
\cite{Wei,Hassard} is required, numerical bifurcation analysis can
be used to investigate it. In Fig. \ref{fig:fig3.eps}, we obtain
the variation of the amplitude $\Delta{x_1}=\mathrm{max}\big(x_1
(t)\big)-\mathrm{min}\big(x_1 (t)\big)$ for the limit cycle branch
emerging from the Hopf bifurcation point `1' with increasing $\tau$
at a fixed $\beta_1 = 7.2$. The solid and the dashed curves
represent the stable and the unstable branches, respectively. The result
shows that a smooth transition from the steady state to the limit cycle
arises across the Hopf point, supporting the Hopf
bifurcation being supercritical. The limit cycle branch arising from
the Hopf point loses its stability at the period-doubling
bifurcation point `2'.

In Fig. \ref{fig:fig4.eps}, we show the time series and the phase
plot of the model in Eq. (\ref{eq:23}) when it exhibits chaotic
behavior ($\lambda_{1} \simeq 0.038$). The results tell us that the
love dynamics of two individuals exhibits stormy patterns of
feelings and a long-time unpredictable state.

\begin{figure}
\begin{center}
\includegraphics[width=8.8cm]{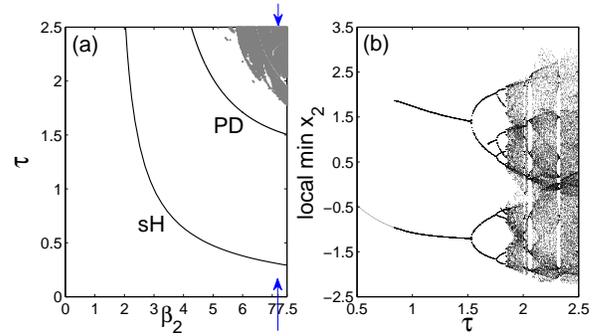}
\caption{(a) Bifurcation diagram in $(\beta_2 ,\tau)$ for the parameter
value: $\alpha_1 = \alpha_2 = 1$, $\beta_1 = 1$, and $\gamma_1 A_2
= \gamma_2 A_1 = 0.5$. (b) Orbit diagram of local minima of $x_2$ at
a fixed value $\beta_{2}=7.2$.}\label{fig:fig5.eps}
\end{center}
\end{figure}

For the parameter space $(\beta_2 ,\tau)$, we show the bifurcation
diagram in Fig. \ref{fig:fig5.eps}(a). It consists of sH and PD
curves including a chaotic region. In Fig. \ref{fig:fig5.eps}(b),
we plot the orbit diagram of local minima of $x_2$ for fixed
$\beta_{2} = 7.2$. Here, the orbit diagram is obtained from
$x_{1}=x_{2}=0$ for $t\leq0$ and clearly shows a period-doubling
route to chaos. The results of Figs. \ref{fig:fig5.eps}(a) and
\ref{fig:fig5.eps}(b) are same as those of Figs. \ref{fig:fig2.eps}(a) and
\ref{fig:fig2.eps}(b) except for the bistable phenomena. In both
parameter spaces $(\beta_{1},\tau)$ and $(\beta_{2},\tau)$, we
observe the same bifurcation structure consisting of a supercritical
Hopf and a cascade of period-doubling bifurcations, that is, a
period-doubling route to chaos.

Through a numerical bifurcation analysis, we have confirmed the
theoretical predictions on the occurrence of a Hopf bifurcation and
obtained the following results: For $\beta_i$ smaller than
${\beta_i}^c$, a steady state of love dynamics cannot be
destabilized by a Hopf bifurcation, no matter how long $\tau$ is. That
is, the existence of time delay does not influence a plateau of the love
affair for insensitive couple. ${\beta_i}^c$ is a minimum value of
$\beta_i$, which satisfies the inequality of Eq. (\ref{eq:17}) for given
parameter values. In those cases, ${\beta_1}^c \simeq 1.27$ and
${\beta_2}^c \simeq 1.68$. For $\beta_i > {\beta_i}^c$, the
parameter value of $\tau$, where the supercritical Hopf bifurcation
arises, becomes shorter as $\beta_i$ increases. For large $\beta_i$
and long $\tau$, the limit cycle undergoes a cascade of period
doubling bifurcations resulting in chaotic motion, i.e., stormy
patterns of feelings.

\subsection{Synergic Couple}
\label{sec:4-2}

Let us now investigate the synergic couple. For the same reason
as in the non-synergic case, the couple composed of a secure and a
non-secure individuals are considered. Then, the love dynamics is
given by

\begin{eqnarray}\label{eq:24}
\dot{x_1}(t) = -\alpha_1 x_{1}(t) &+&
{R_1}^s \big(x_{2}(t-\tau)\big) \\
&+&\big\{ 1+S_{1}\big(x_{1}(t)\big) \big\}\,\gamma_1 A_{2}, \nonumber \\
\dot{x_2}(t) = -\alpha_2 x_{2}(t) &+& {R_2}^n
\big(x_{1}(t-\tau)\big) \nonumber \\
&+&\big\{ 1+S_{2}\big(x_{2}(t)\big) \big\}\,\gamma_2 A_{1},
\nonumber
\end{eqnarray}

\noindent where the functional forms of the secure return ${R_1}^s$
and the non-secure return ${R_2}^n$ are described by Eqs.
(\ref{eq:4}) and (\ref{eq:5}), respectively. Here, the synergic
functions ${S_1}$ and ${S_2}$ are given by Eq. (\ref{eq:7}).

\begin{figure}
\begin{center}
\includegraphics[width=8.8cm]{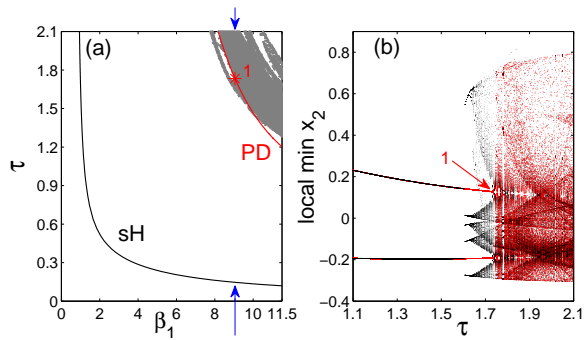}
\caption{(a) Bifurcation diagram in $(\beta_1 ,\tau)$ for the parameter
value: $\alpha_1 = \alpha_2 = 1$, $\beta_2 = 1$, $\gamma_1 A_2 =
\gamma_2 A_1 = 0.5$, and $\sigma_1 = \sigma_2 = 0.5$. (b) Orbit
diagram of local minima of $x_2$ at a fixed value
$\beta_{1}=9.05$.}\label{fig:fig6.eps}
\end{center}
\end{figure}

\begin{figure}
\begin{center}
\includegraphics[width=8.8cm]{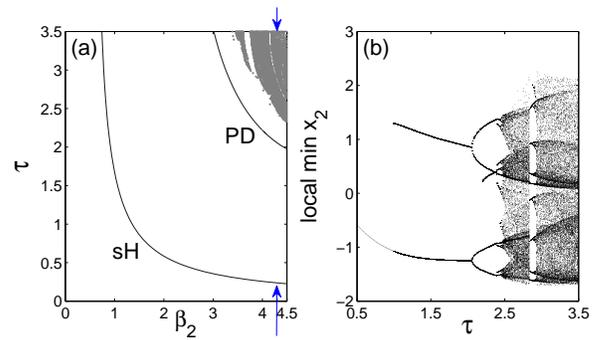}
\caption{(a) Bifurcation diagram in $(\beta_2 ,\tau)$ for the parameter
value: $\alpha_1 = \alpha_2 = 1$, $\beta_1 = 1$, $\gamma_1 A_2 =
\gamma_2 A_1 = 0.5$, and $\sigma_1 = \sigma_2 = 0.5$. (b) Orbit
diagram of local minima of $x_2$ at a fixed value
$\beta_{2}=4.3$.}\label{fig:fig7.eps}
\end{center}
\end{figure}

For the parameter space $(\beta_{1},\tau)$, we plot the bifurcation
and the orbit diagrams of the model in Eq. (\ref{eq:24}) in Figs.
\ref{fig:fig6.eps}(a) and \ref{fig:fig6.eps}(b), respectively. In
Fig. \ref{fig:fig6.eps}(b), the black and the red dots are obtained from
$x_{1}(t)=x_{2}(t)=0$ and $x_{1}(t)=2.4,x_{2}(t)=0.2$ for $t \leq
0$, respectively. The above results exhibit bistable phenomena. The
point `1' in Figs. \ref{fig:fig6.eps}(a) and \ref{fig:fig6.eps}(b)
represents the same point, which corresponds to a period doubling
of the limit cycle branch arising from the sH curve. However, we cannot
continue other bifurcation curves involved in the bistable
phenomena, such as the LPC and the black PD curves in Fig.
\ref{fig:fig2.eps}(a).

In Figs. \ref{fig:fig7.eps}(a) and \ref{fig:fig7.eps}(b), we present
the results of the numerical bifurcation analysis for the parameter
space $(\beta_{2},\tau)$. Here, the orbit diagram is obtained from
$x_{1}(t)=x_{2}(t)=0$ for $t \leq 0$. Equivalently to the
non-synergic case, the results of Figs. \ref{fig:fig7.eps}(a) and
\ref{fig:fig7.eps}(b) are the same as those for Figs. \ref{fig:fig6.eps}(a) and
\ref{fig:fig6.eps}(b) except for the bistable phenomena. Therefore,
we explicitly show that a supercritical Hopf and a cascade of period-doubling 
bifurcations, i.e., a period-doubling route to chaos, is a
universal bifurcation structure in our models in Eqs. (\ref{eq:23}) and
(\ref{eq:24}). For the synergic case, we identify that ${\beta_1}^c
\simeq 0.79$ and ${\beta_2}^c \simeq 0.60$.

\section{Conclusion}
\label{sec:5}

In conclusion, we have investigated the effect of time delay on
a simplified mathematical model that describes the dynamics of love
between two individuals in the field of social science. By analyzing
the characteristic equation of linearization of the model, we have
theoretically proven that if no one exhibits a non-secure return in
both cases of synergic and non-synergic couples, then the existence
of time delay cannot disturb a steady state of love dynamics.
However, if at least one of them has a non-secure return, then the
time delay on the return function can cause a Hopf bifurcation and a
cyclic love dynamics. Through a numerical bifurcation analysis, we
have confirmed the theoretical predictions on the occurrence of the Hopf
bifurcation and obtained the universal bifurcation structure
consisting of a supercritical Hopf and a cascade of period-doubling
bifurcations, resulting in chaotic motion, which exhibits stormy
patterns of feelings and a long-time unpredictable state. We have
also ascertained that the existence of time delay does not
influence a steady state of the love affair for an insensitive couple.

Through further investigation, we hope that the time delay effect on
a more realistic love dynamical model, including a love triangle, can be
investigated. Moreover, it seems that such modeling approach with
time delay can be applied to various dynamical phenomena in the
field of social science, including a study of
the proper relation between supplier and consumer in the field of management.

\begin{acknowledgements} This research was supported by World Class
University program funded by the Ministry of Education, Science and
Technology through the National Research Foundation of
Korea (R31-20002). \end{acknowledgements}

\end{document}